# MoO$_2$/C composites prepared by tartaric acid and glucose-assisted sol-gel processes as anode materials for Lithium-ion batteries

G.S. Zakharova[1*], L. Singer[2*], Z.A. Fattakhova[1], S. Wegener[2], E. Thauer[2], Q. Zhu[3], E.V. Shalaeva[1], R. Klingeler[2,4]

[1]*Institute of Solid State Chemistry, Ural Division, Russian Academy of Sciences, Yekaterinburg, Russia*

[2]*Kirchhoff Institute of Physics, Heidelberg University, Heidelberg, Germany*

[3]*Institute of Material Science and Engineering, Wuhan University of Technology, Wuhan, PR China*

[4]*Centre for Advanced Materials (CAM), Heidelberg University, Heidelberg, Germany*

*Both authors contributed equally.

**Abstract**

MoO$_2$/C-composites have been fabricated for the first time by a tartaric acid/glucose-assisted sol-gel method with post-annealing at 500 ºC for 1 h in N$_2$ flow. The synthesized materials were fully characterized with respect to structure, morphology, and electrochemical properties. Compared with tartaric acid-assisted products, the adoption of glucose as carbon source effectively increases the carbon content in the composites. Irrespective of the organic component, the composites exhibit low crystallinity and small grain size. This results in good electrochemical performance of the anode materials as confirmed for the glucose-assisted materials which after additional post treatment delivers a competitive electrochemical capacity.

## 1. Introduction

Molybdenum dioxide ($MoO_2$), as an important semiconductor, has great potential applications in the production of chemical sensors [1], catalysts [2], field emission devices [3,4], and solar cells [5]. Additionally, $MoO_2$ has attracted considerable attention as electrode material for lithium ion batteries (LIB) because of its relatively large theoretical capacity (838 mA h g$^{-1}$) [6] and its high metallic conductivity ($6.04 \cdot 10^3$ S cm$^{-1}$ for the individual rods) [7]. However, several of its intrinsic material properties limit the practical application of bulk $MoO_2$, namely strong capacity fading and poor cycling stability originated by the huge volume changes during the charge/discharge processes as well as poor rate performance caused by the slow kinetics [8,11]. A variety of strategies have been employed to overcome this issue such as preparation of nanosized materials with different morphologies [9], doping with nitrogen [10], copper [11], and design of 3D hierarchically structures nanomaterials providing shorter lithium ion/electron diffusion distances and accommodating strain associated with strong volume changes upon cycling [12,13]. Additionally, incorporating $MoO_2$ with functional conductive carbonaceous materials like hollow carbon spheres [14], graphene [15], graphite oxide [16], is regarded as an effective strategy. A further way which advantageously yields homogeneous distribution of the carbon component, thereby increasing cycling stability of the materials is to introduce the carbonaceous phase during calcination of many different organic precursors such as aniline [17], dopamine [18,19], polyethylene glycol [20], alginate [21], ascorbic acid [22], glucose [23], or sucrose [24]. All functionalized materials display enhanced Li-retention performance in comparison with bulk $MoO_2$, based on their higher surface areas, more actives sites, shorter ion diffusion paths, and functioning carbon buffer which accommodate the strain.

Various techniques, such as a hydrothermal method [25,26], thermolysis [27], and electrospinning [28], have been developed to prepare composites based on $MoO_2$. Herein, we present a new sol-gel synthesis approach to obtain $MoO_2$/C composites. This method provides a molecular level-mixing of the starting materials and leads to better chemical homogeneity of the final products. Carbon coated $MoO_2$ particles have been produced by using glucose or tartaric acid as both reducing agents and carbon sources and the electrochemical performance of the resulting $MoO_2$/C composites as electrodes for Li-ion batteries was investigated.

## 2. Experimental

Molybdenum powder Mo (99.95% metal, Alfa Aesar), hydrogen peroxide $H_2O_2$ (30%, Merck), tartaric acid $C_4H_6O_6$ (AppliChem), and glucose $C_6H_{12}O_6$ (Sigma-Aldrich) were utilized for the synthesis. Firstly, 1.0 g of Mo powder was dissolved in 35 ml $H_2O_2$ at 10-15 °C to form a clear yellow solution of peroxomolybdic acid. Secondly, tartaric acid was dissolved in distilled water and stirred for 10 min. The molar ratio of tartaric acid to molybdenum metal was 1 : 1. Then, the solution of tartaric acid was slowly added into the solution of peroxomolybdic acid. To get a gel, the solution was heated at 60 °C under continuous stirring. The resulting gel was dried at 60 °C in air and calcinated at 500 °C for 1 h in $N_2$ flow to yield the $MoO_2$/C composite. At lower annealing temperatures, the materials were amorphous while annealing at 500 °C yields crystalline samples. The as-prepared product is labeled as $MoO_2$/C-T. For comparison, $MoO_2$/C was produced in a glucose-assisted process at a molar ratio of Mo : glucose = 1 : 1 and denoted as $MoO_2$/C-G. To study the effect of pounding on the electrochemical activity, $MoO_2$/C-G powder was colloidally grinded (300 Us$^{-1}$) in ethanol using Ø 1mm $ZrO_2$ balls in a PM 100 planetary mill (Retsch) for 6 h, later annealed[1] at 500 °C for 1 h in $N_2$ and afterwards hand grinded in a mortar for 15 min; this sample is labeled as $MoO_2$/C-G(M).

X-ray diffraction (XRD) patterns were obtained on a Bruker AXS D8 Advance Eco using Cu Kα radiation with a step size of Δ2θ = 0.02°. The morphology of the powder was determined by a ZEISS Leo 1530 and a JEOL

---
[1]

JSM-7610F scanning electron microscope (SEM), a JEOL JEM 2100 and a JEMe200 CX transmission electron microscope (TEM), respectively. In order to monitor the microstructural changes of $MoO_2$ upon cycling, ex-situ SEM studies were performed using a JEOL JSM-7610F scanning electron microscope. The cycled electrodes were disassembled in an Argon glove box, washed with ethylene carbonate and then dried overnight.

Fourier transform infrared (FT-IR) spectra were recorded using Spectrum One B (Perkin-Elmer) with an automatic diffuse reflectance accessory. A thoroughly ground sample was applied as a thin layer to a purpose-designed holder plate. Thermogravimetric analysis (TG-DSC-MS) with a heating rate of 10 K·min$^{-1}$ starting from room temperature up to 750 °C under flowing air was carried out using STA 449 $F_3$ Jupiter thermoanalyzer (Netzsch) coupled with a QMS 403 mass spectrometer. Raman spectra were measured with a Renishaw U1000 spectrometer at a laser wavelength of 532 nm.

Electrochemical measurements were carried out in Swagelok-type cells (see [29]). Working electrodes were prepared by mixing the active material, carbon black and polyvinylidene fluoride (PVDF) in N-methyl-2-pyrrolidinone (NMP) and stirred for 24 h before the resulting slurry was pasted on copper net current collectors. The as-prepared electrodes were dried in a vacuum oven (80 °C, 10 mbar) overnight, pressed and dried again. Fiberglass (Whatman GF/D) was used as separator and pure lithium metal foil (Aldrich) as counter electrode. The electrolyte was 1 M $LiPF_6$ in a mixture of ethylene carbonate and diethylcarbonate (1 : 1 by weight). Cell assembly was carried out in an Ar-filled glovebox with controlled moisture and oxygen concentration. Cyclic voltammetry (CV) at a scan rate of 0.1 mV s$^{-1}$ and galvanostatic cycling with potential limitation (GCPL) at specific currents of 100 mA g$^{-1}$, both in the voltage range of 0.01 - 3.0 V vs. Li/Li$^+$, were carried out on a VMP3 potentiostat (BioLogic) at 25 °C.

## 3. Results and discussion

Fig. 1 shows X-ray diffractograms of the synthesized materials, i.e., $MoO_2$/C-G, $MoO_2$/C-G(M), and $MoO_2$/C-T. All observed peaks correspond to monoclinic $MoO_2$ and are accordingly indexed in a monoclinic lattice system with space group $P2_1/c$. The post treatment of the as-prepared composite does not change its structure. The experimentally determined lattice parameters of $MoO_2$ in the composite $MoO_2$/C materials are in a good agreement with theoretical values from JCSD no. 72-4534 (cf. Table 1).

Table 1. Experimentally determined lattice parameters of $MoO_2$ in $MoO_2$/C-G, $MoO_2$/C-G(M), and $MoO_2$/C-T as well as values from JCSD no. 72-4534.

|  | $a$ (Å) | $b$ (Å) | $c$ (Å) | $\beta$ (deg.) | $V$ (Å$^3$) |
|---|---|---|---|---|---|
| $MoO_2$/C-G | 5.652(6) | 4.832(5) | 5.594(7) | 120.39(6) | 131.7(3) |
| $MoO_2$/C-G(M) | 5.679(5) | 4.808(9) | 5.619(8) | 120.48(2) | 132.2(3) |
| $MoO_2$/C-T | 5.648(3) | 4.912(6) | 5.609(1) | 119.24(7) | 135.8(2) |
| $MoO_2$ | 5.6109 | 4.8562 | 5.6285 | 120.95 | 131.53 |

The XRD data exhibit broad diffraction peaks with rather weak intensity which suggests that $MoO_2$ formed in the composites displays poor crystallinity and small crystal size. The mean primary crystallite size of the as-synthesized samples can be estimated using Scherrer's equation [30]:

$$D = K\lambda/\Delta(2\theta)\cos\theta, \qquad (1)$$

where $D$ is the average grain size based on the particular reflecting crystal face (hkl) direction, $K$ is a shape factor which can be approximated to 0.9, $\lambda$ is the wavelength of the applied Cu K$\alpha$ radiation, $\Delta(2\theta)$ is the full width at half-maximum of the diffraction peak, and $\theta$ is the Bragg angle. Analysis[2] of the (011), (200), and (022) peaks yield to a mean primary crystallite size of around 6 nm for MoO$_2$/C-G, MoO$_2$/C-G(M), and MoO$_2$/C-T composites under study. The crystal structure of monoclinic MoO$_2$ is depicted in Fig. 1. It consists of MoO$_6$ octahedrons which form channels along the *a*-axis which are supposed to facilitate Li$^+$ transport during electrochemical cycling.

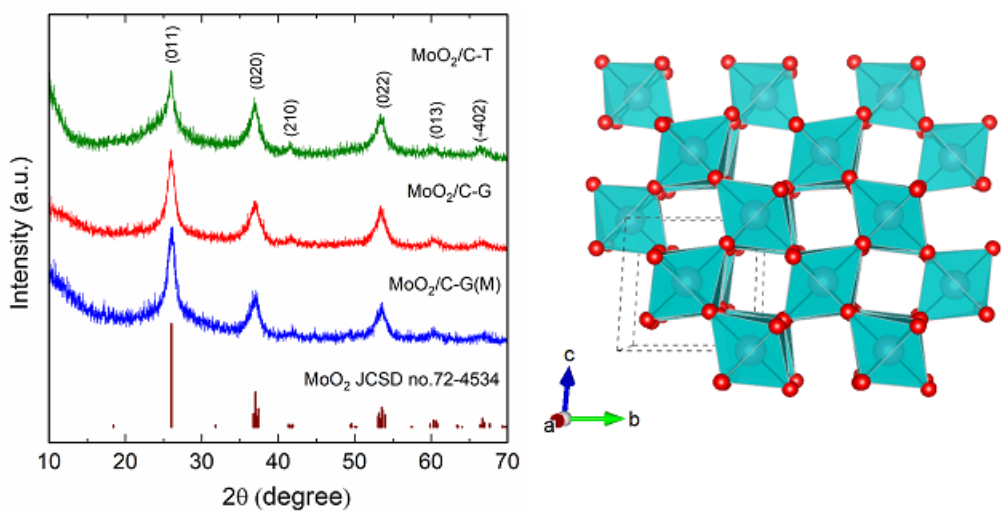

Fig. 1. XRD patterns of the MoO$_2$/C-G and the MoO$_2$/C-T composites (left), and a schematic of the corresponding crystal structure of MoO$_2$ (right).

Scanning electron microscopy images in Fig. 2 show the morphology of the as-prepared composites. Both as-prepared composite materials MoO$_2$/C-T and -G consist of irregular micro-sized agglomerates (Fig. 2a, c). The underlying primary particles, which are well visible in the high-magnification SEM images in Fig. 2b and d, in both composites consist of round grains with an average particle size of less than 20 nm A further post treatment of the sample results in a visible size reduction of the agglomerates (Fig. 2e-f).

---
[2] Additional contributions to peak broadening are neglected.

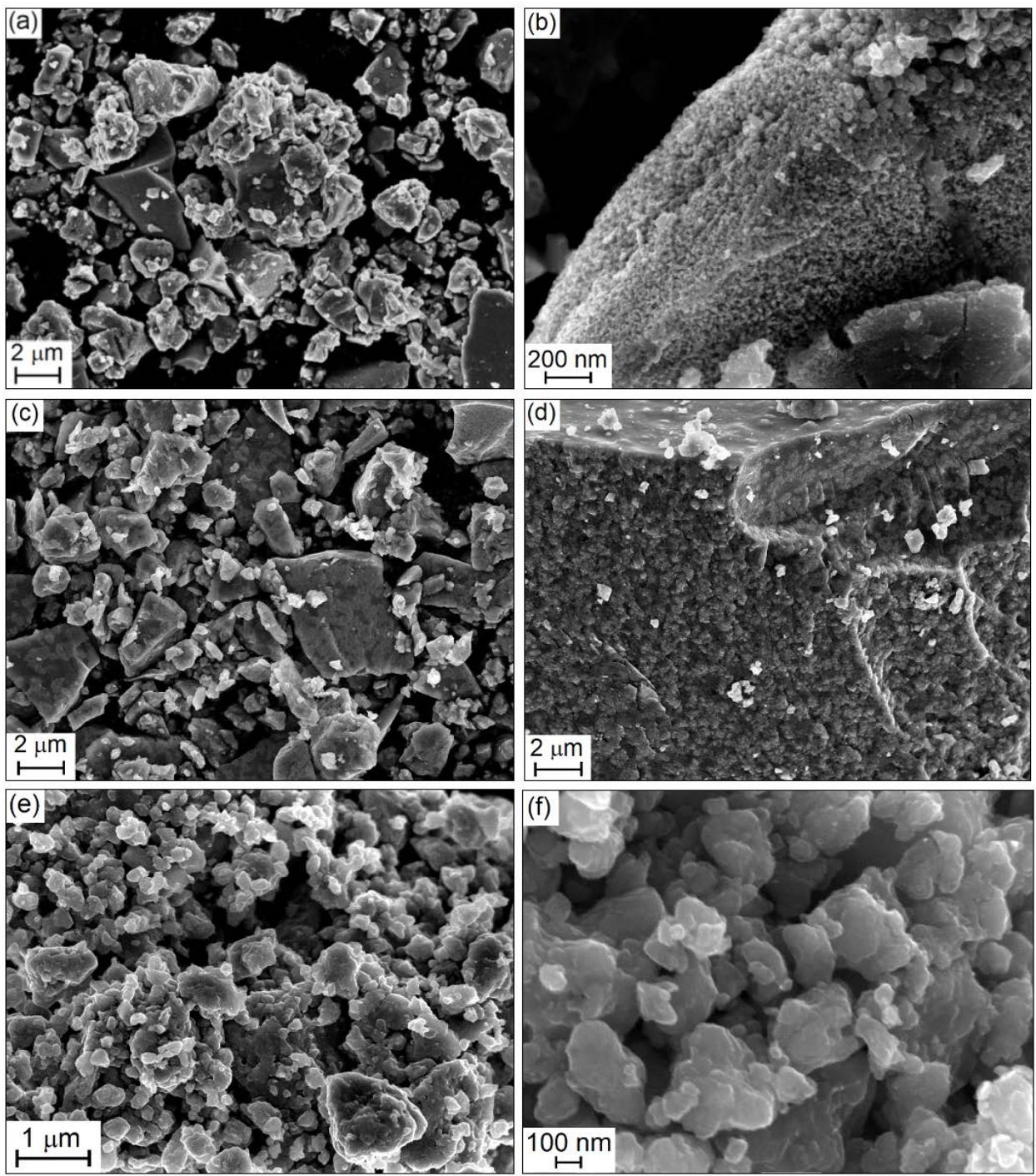

Fig. 2. SEM and high-magnification SEM images of (a, b) MoO$_2$/C-T, (c, d) MoO$_2$/C-G, and (e, f) MoO$_2$/C-G(M) obtained by a post treatment (see the text).

Additional detailed information on MoO$_2$/C-G is obtained from the TEM images in Fig. 3a. The images show the presence of MoO$_2$/C-G nanocrystallites forming large agglomerates that are dispersed in an amorphous carbon matrix.

Lattice fringes visible in the high-resolution TEM (HR-TEM) image (Fig. 3b) indicate lattice display spacings of approximately 0.35 nm. This value corresponds well with the (011) plane of $MoO_2$ which according to our XRD analysis has a plane distance of 0.3417(3) nm in $MoO_2$/C-G. Fig. 3b also shows carbon layers of around 10 nm in thickness around the crystallites. Selected area electron diffraction (SAED) confirms the crystalline structure of $MoO_2$ in the $MoO_2$/C-G nanocomposite (Fig. 3c). Debye rings from the SAED pattern demonstrate the ultrafine structure of crystalline $MoO_2$ without any preferred orientation of the individual crystallites. The Debye rings can be attributed to the monoclinic phase of $MoO_2$ with $P2_1/c$ space group symmetry, which agrees with the XRD results.

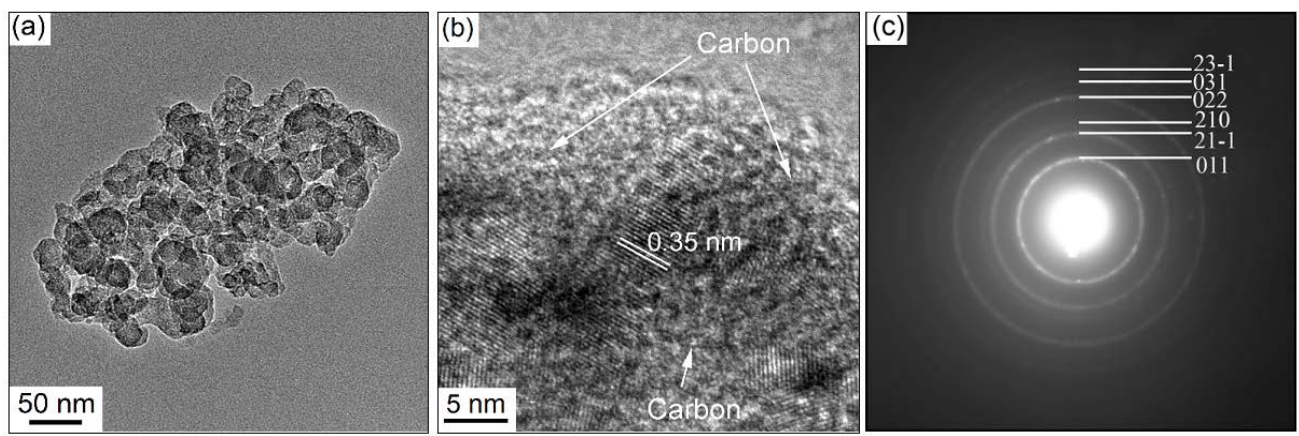

Fig. 3. (a) TEM image, (b) HRTEM image, and (c) SAED pattern of $MoO_2$/C-G composite.

Raman spectra of the $MoO_2$/C composites shown in Fig. 4a provide information on both the $MoO_2$ and the carbon components of the nanocomposite as the observed Raman peaks refer to Mo-O as well as carbon modes. The prominent carbon modes are the G-band which corresponds to the $sp^2$-bonded carbon atoms and the D-band associated with defects or disorder mainly due to $sp^3$-hybridization of carbon atoms [31]. The D- and G-bands of $MoO_2$/C-G composite arise at 1372 $cm^{-1}$ and 1586 $cm^{-1}$, respectively. In comparison, the G- and D-bands of $MoO_2$/C-T are at slightly lower wavenumbers of 1365 and 1575 $cm^{-1}$. The ratio of the maximum intensities of these peaks, $I_D/I_G$, was calculated as 0.78 and 0.71 for $MoO_2$/C-G and $MoO_2$/C-T composites, respectively. This indicates that in comparison with tartaric acid, glucose promotes the formation of defects and disorder in the carbon component of the composite. In comparison with the pure $MoO_2$ phase [32], a richer Raman spectrum of $MoO_2$/C composites with intensive bands below 1000 cm$^{-1}$ resulting from different vibration modes of Mo-O is observed. The presence of the peaks located at 819 and 993 cm$^{-1}$ corresponding to the stretching vibrations of Mo=O bonds indicates the formation of the layered structure of the compounds similar to $MoO_3$ [33]. This phenomenon has been reported by Camacho-Lopez *et al.* [34] who associated it with the oxidation of $MoO_2$ to $MoO_{2+\delta}$ by laser irradiation.

FT-IR spectra of the $MoO_2$/C composites are shown in Fig. 4b. There are two typical vibration modes belonging to the $MoO_2$ phase [35]. The stretching vibrations of Mo=O bonds are demonstrated by the characteristic bands at 957 and 955 cm$^{-1}$ for $MoO_2$/C-T and $MoO_2$/C-G composites, respectively. The bridge stretching vibration of Mo-O-Mo bonds are displayed by the bands at 718 and 693 cm$^{-1}$ for $MoO_2$/C-T and $MoO_2$/C-G composites, respectively. The peaks at 1598 cm$^{-1}$ ($MoO_2$/C-T) and 1600 cm$^{-1}$ ($MoO_2$/C-G) are assigned to the C-O stretching vibration in the carboxylic group which may result from incomplete decomposition of the organic groups in the carbon component. The weak bands located at around 3400 cm$^{-1}$ and 1640 cm$^{-1}$ are associated with the stretching and bending vibrations of O–H, respectively, originating from trace amounts of adsorbed water in the powder. It is worth noting that the peaks which are attributed to pure glucose [36] and tartaric acid [37] are not part of the respective spectra.

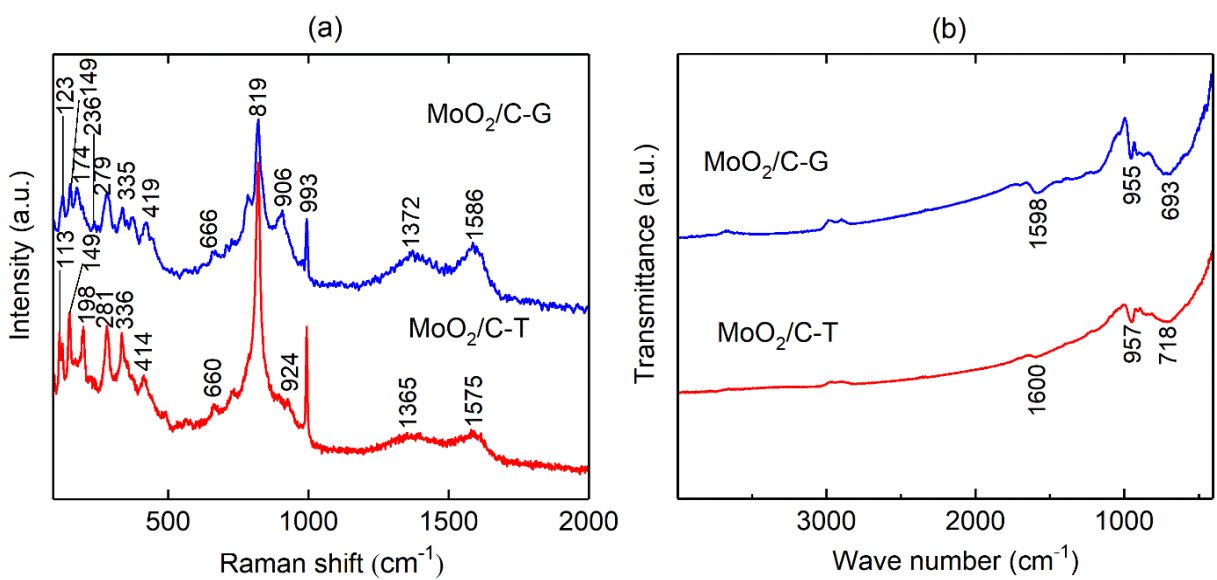

Fig. 4. (a) Raman and (b) FT-IR spectra of MoO$_2$/C-G and MoO$_2$/C-T composites.

In order to determine the thermal stability of MoO$_2$/C composites in air and to quantify the amount of carbon within the materials, TG-DSC-MS studies of as-prepared materials were carried out (Fig. 5). The TG curve of the MoO$_2$/C-G composite evidences two steps of weight loss (Fig. 5a). We attribute the first weight loss of 5.7 wt% in the temperature regime from 70 ºC to 300 ºC mainly to evaporation of water. It appears as a weak and broad endothermic process centered at 115 ºC. The second weight loss of 23.7 wt% from 300 ºC to 690 ºC is caused by oxidation of carbon components and the release of CO$_2$ gases. The mass spectrometry curve (ion current versus temperature) indicates that the main gaseous product of MoO$_2$/C-G decomposition is CO$_2$ (m/z = 44 a.e.m.). However, the carbon component of MoO$_2$/C-G was not completely oxidized to CO$_2$ at 690 ºC. Despite the appearance of a two-step process somehow similar to what is observed in MoO$_2$/C-G, the TG curve of MoO$_2$/C-T shown in Fig. 5b shows several differences. Again, a first stage from 25 to 300 ºC with corresponding weight loss of 3.6 wt% can be assigned to evaporation of surface-adsorbed water. The second region of mass loss (2.4 wt%) from 300 to 514 ºC is accompanied by a broad exothermal peak. It is attributed to the decomposition and full oxidation of organics. From the mass spectrometry curve, it can be revealed that the main gaseous product of MoO$_2$/C-T decomposition is CO$_2$ (m/z = 44 a.e.m.). Additionally, the three steps observed in TGA suggest that the carbon component in the MoO$_2$/C-T composite has three states differently bound to the main phase (MoO$_2$). At around 550 °C, the TG data indicate an increase of sample weight by 2.9 wt%. This process is attributed to the oxidation of Mo$^{4+}$ to Mo$^{6+}$ and results in the formation of MoO$_3$ as a final thermolysis product. Based on the assumptions made above, the calculated carbon contents in MoO$_2$/C-T and MoO$_2$/C-G composites are 2.4 and more than 23.7 wt%, respectively.

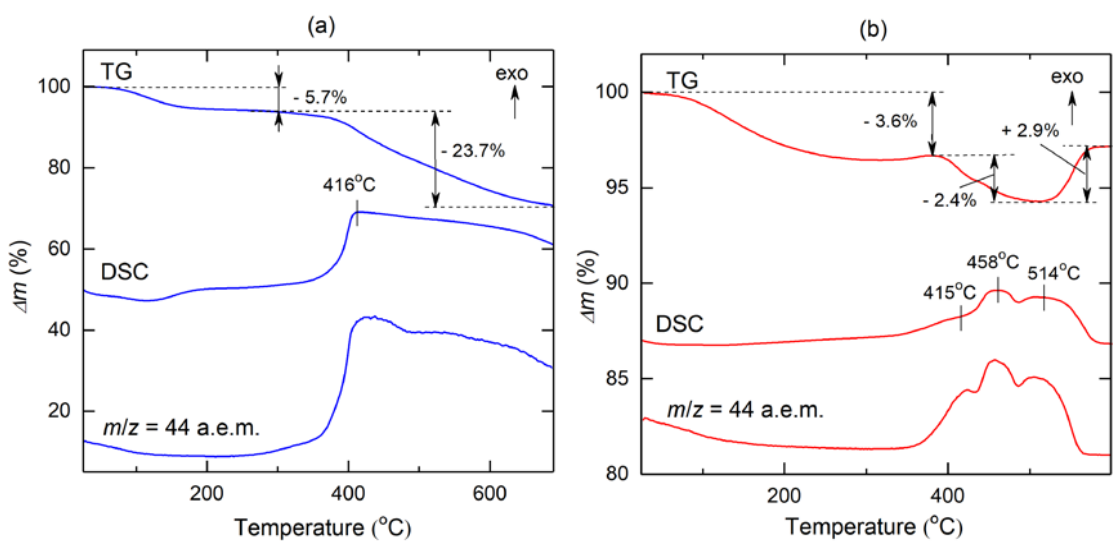

Fig. 5. Thermogravimety, DSC, and mass spectroscopy curves of (a) MoO$_2$/C-G and (b) MoO$_2$/C-T composites.

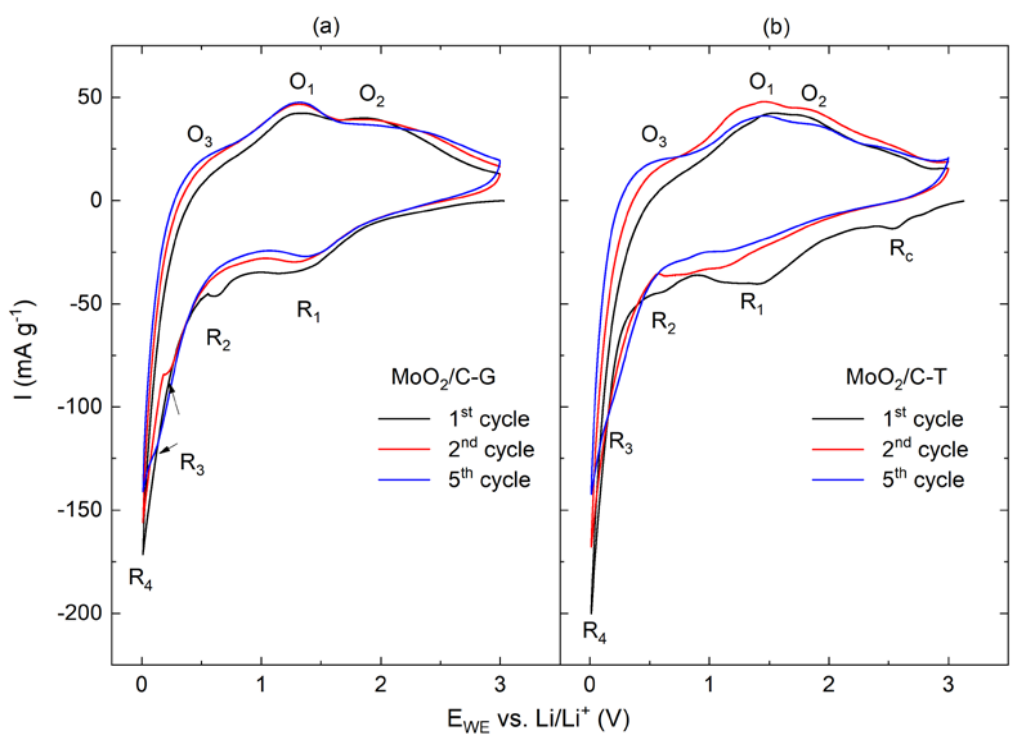

Fig. 6. Cyclic voltammograms of (a) MoO$_2$/C-G and (b) MoO$_2$/C-T composites recorded at a scan rate of 0.1 mV s$^{-1}$.

The electrochemical properties of the composite materials obtained as described are investigated with respect to their applicability as electrode materials in LIB by means of CV and GCPL. Fig. 6 displays the first, second, fifth CV curve of MoO$_2$/C-G and MoO$_2$/C-T. The reversible reactions of MoO$_2$ with lithium are described by the following electrochemical reactions [16,38]:

$$MoO_2 + xLi^+ + xe^- \leftrightarrow Li_xMoO_2, \quad (0 \leq x \leq 0.98) \quad (2)$$

$$Li_xMoO_2 + (4-x)Li^+ + (4-x)e^- \leftrightarrow Mo + 2Li_2O. \quad (3)$$

Due to the deliberate low crystallinity of the samples, mainly extended areas of electrochemical activity instead of well-defined peaks are visible in the CV. In the first reductive sweep of MoO$_2$/C-G, four different cathodic features are observed. The elevation at around 1.25 V (R1) can be attributed to the phase transition between the orthorhombic and the monoclinic phase upon lithium insertion (Eq. 2) as suggested by Dahn et al. [39]. Due to the absence of the peak R2 at around 0.6 V in further cycles, it is most likely assigned to the solid electrolyte interface (SEI) formation which is consistent with reports by Luo *et al.* [40] and Xu *et al.* [41]. The reversible conversion reaction (Eq. 3) can be assigned to the reduction peak R3, parts of R4 and the elevation O3 [38]. The two overlapping oxidation peaks O2 and O1 can be ascribed to the phase transitions (monoclinic-orthorhombic-monoclinic) in partially lithiated $Li_xMoO_2$ [42]. The main difference between the CV of $MoO_2$/C-G (Fig. 6a) and $MoO_2$/C-T (Fig. 6b) is the presence of the feature Rc at around 2.5V in the first reductive sweep of the latter. In the literature, Rc is sometimes observed [27,43] and sometimes not [42,44] and its origin is yet unclear.

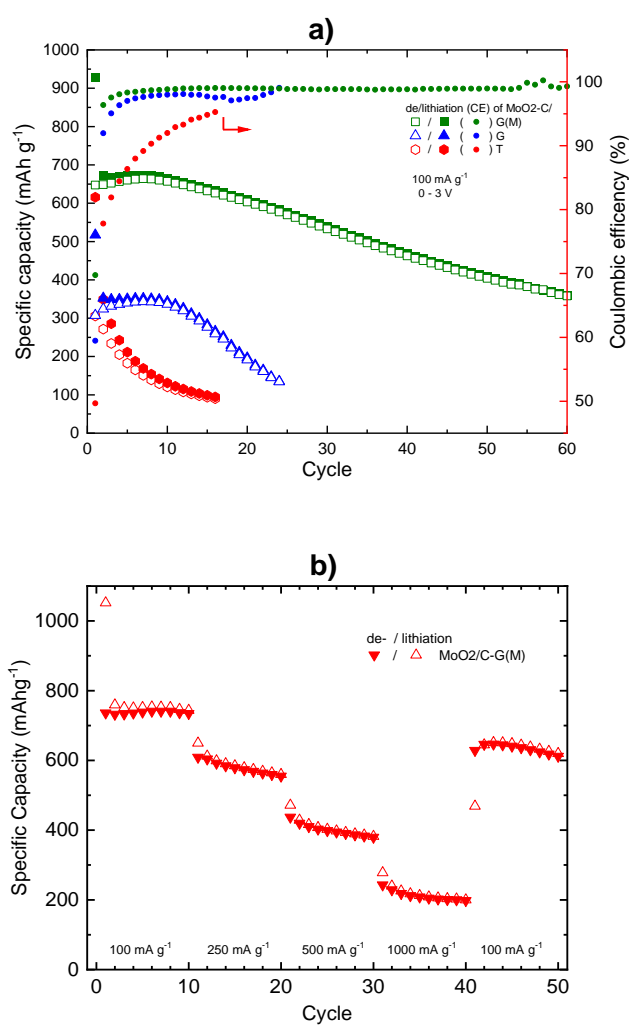

Fig. 7 a) Specific charge/discharge capacities and associated coulombic efficiencies of $MoO_2$/C electrodes and b) rate capability test with different cycling rates of $MoO_2$/C-G(M) electrode measured by GCPL between 0.01 V and 3 V.

The cycling performance of as-prepared $MoO_2$/C-G- and $MoO_2$/C-T- as well as of $MoO_2$/C-G(M)-based electrodes is presented in Fig. 7a. The corresponding potential profiles are shown in Fig. S1-S3. The electrochemical performance of the $MoO_2$/C composites show strong differences. The highest discharge capacity in the first and twentieth cycle of 928 mAh g$^{-1}$ and 610 mAh g$^{-1}$ is reached by the post treated sample $MoO_2$/C-G(M). In contrast, the as-prepared material $MoO_2$/C-G achieves 516/195 mAh g$^{-1}$, respectively. Coloumbic efficiencies also shown in Fig. 7a confirm significant irreversible effects not only for $MoO_2$/C-T but also for $MoO_2$-/C-G. All materials display reduced Coulombic efficiency in the first cycles which is attributed to common irreversible processes as electrolyte decomposition and formation of the SEI [22,40,41]. Similarly to what is concluded from the evolution of capacity upon cycling, Coulombic efficiencies for $MoO_2$/C-G(M) are highest and amount to about 98 %. $MoO_2$/C-T has in comparison to $MoO_2$/C-G a high initial discharge capacity of 617 mAh g$^{-1}$ but shows a much severe drop in capacity already in the second cycle. We attribute this to the much lower (i.e., by more than 20%) carbon content. Such a low carbon content may result in an increased contact loss between the active material and the current collector during cycling. This degeneration process is well known for conversion-type materials and occurs due to the volume expansion and the associated induced pulverization during the conversion reaction [45–47]. Comparison of the treated and untreated $MoO_2$/C-G samples shows that the post treated sample $MoO_2$/C-G(M), which features a reduced size of the agglomerates (Fig. 2e-f), achieves nearly 2 times the capacity of $MoO_2$/C-G over all cycles. As expected, the size reduction of the agglomerates by the post treatment and the associated surface enlargement leads to an increased electrochemical activity of $MoO_2$/C-G as electrolyte diffusion is facilitated [48]. The initial increase in capacity of the $MoO_2$/C-G samples may be attributed to a gradual decomposition of $Li_xMoO_2$ in the conversion reaction during cycling [43], the formation of crystalline defects which is a common phenomenon in oxide anodes resulting in increasing capacity [49][50*], and/or a reaction including the SEI [50]. The rate capability of $MoO_2$/C-G(M) at cycle rates between

100 and 1000 mA g$^{-1}$ displayed in Fig. 7b implies a reversible capacity of 740, 580, 400, and 210 mA h g$^{-1}$, at a current density of 100, 250, 50 and 1000 mA g$^{-1}$, respectively. The reversible capacitance reaches 650 mA h g$^{-1}$ when the current density is set again to 100 mA g$^{-1}$, which proves good reversibility of the electrode. To characterize the microstructure details of the MoO$_2$/C-G(M) based electrode before and after galvanostatic cycling, *ex-situ* SEM measurements were performed (Fig. S4). The comparison of the SEM images before (Fig. S4a) and after (Fig. S4b) cycling reveals that the individual particles change their surface texture, which is likely due to the known degeneration mechanisms such as SEI formation [51] and/or agglomeration. Pulverization as a possible degeneration phenomenon for MoO$_2$/C-G(M) seems to be less critical as the high-resolution SEM images show that even after 60 cycles the binder network is intact and no clear cracks are visible.

To relate the electrochemical performance of the here presented MoO$_2$/C-G(M) composite to the literature, Table 2 lists the discharge capacities of various MoO$_2$/C composite anodes. The MoO$_2$/C-G(M)-based electrode studied at hand exhibits a superior electrochemical capacity compared to the materials obtained by Che *et al.* [21], Yoon *et al.* [52] and Luo *et al.* [40] and is slightly better than the materials reported by Sun *et al* [53] and Hu *et al.* [54]. However, MoO$_2$/graphene oxide composite materials generally show higher capacities [16,41,55]. In contrast to these materials, the MoO$_2$/C-G(M) composite reported here benefits from a straightforward, environment-friendly, and well-controllable synthesis process, which is based on cost-effective starting materials. Evidently, the electrochemical performance of MoO$_2$/C composites mainly depends on morphological features as carbon coating, particle size and surface texture which are directly affecting the transport path length of electrons as well as Li-ions.

Table 2. Comparison of the electrochemical cycling performance of $MoO_2$-based composites obtained by various synthesis methods.

| Material | Synthesis method | Current (mA g$^{-1}$) | Discharge capacity (mAh g$^{-1}$)/cycle no* | Ref. |
|---|---|---|---|---|
| $MoO_2$/C-G(M) nanoparticles | Glucose-assisted sol-gel | 100 | 660/10 | this work |
| $MoO_2$/C nanoparticles | Alginate-assisted sol-gel | 200 | 300/10 | [21] |
| Nitrided $MoO_2$ | hydrothermal | 120 | 310/10 | [52] |
| $MoO_2$/C nanofibers | Electrospinning | 100 | 500/10 | [40] |
| $MoO_2$/C nanoparticles | Impregnation of carbon matrix | 100 | 600/10 | [53] |
| $MoO_2$/graphite oxide nanoparticles | Solvothermal | 100 | 620/10 | [54] |
| $MoO_2$/graphite oxide nanoparticles | Solvothermal | 100 | 800/10 | [16] |
| $MoO_2$/C cage-like particle | Hydrothermal reduction | 200 | 800/10 | [22] |
| $MoO_2$/exfoliated graphene oxide | Solid-state graphenothermal reduction | 100 | 850/10 | [55] |
| $MoO_2$/graphene oxide | Decomposition | 100 | 1200/10 | [41] |

*For better comparison, data from the same cycle number are shown as read-off from reported figures.

**Conclusions**

In summary, a tartaric acid and glucose based sol–gel method combined with thermal reduction has been utilized to synthesize hybrid $MoO_2/C$ nanocomposites. This synthetic strategy is simple, cost-effective, and promising for large-scale industrial production of $MoO_2/C$ composites. In the as-prepared composites, $MoO_2$ nanoparticles with a size of about 20 nm are embedded into a ~10 nm thick amorphous carbon matrix, which serves as a buffer layer preventing the degradation of $MoO_2$ nanoparticles during charge/discharge processes.

The observed electrochemical performances of the prepared materials underline the relevance of the used carbon source and the received morphology for the application in LIB. Compared to untreated $MoO_2/C$-G, post treated $MoO_2/C$-G(M) shows a clearly improved capacity. The post treated $MoO_2/C$-G(M) features a remarkably enhanced, competitive specific capacity of 660 mAh $g^{-1}$ at 100 mA $g^{-1}$ in the tenth cycle. Therefore, the results show that this simple and cost-effective synthesis approach may also be successfully applied to other metal oxides.

**Acknowledgements**

This work was supported by the Deutsche Forschungsgemeinschaft through projects KL 1824/12-1 and KL 1824/14-1. G.Z. acknowledges support of the state order via the Ministry of Science and High Education of Russia (Theme no. AAAA-A19-119031890025-9). Partial support by the BMWi through project 03ET6095C (HiKoMat) is acknowledged. The authors thank I. Glass for experimental support.